\documentclass[aps,pra,superscriptaddress,twocolumn,longbibliography,10pt,notitlepage,nofootinbib]{revtex4-2}
\usepackage[utf8]{inputenc}
\usepackage[T1]{fontenc}

\usepackage{amsmath}
\usepackage{amssymb}
\usepackage{amsfonts}
\usepackage{amsthm}
\usepackage{bbm}
\usepackage{bm}
\usepackage{quantikz}

\usepackage{xcolor}
\definecolor{color1}{rgb}{0,0,0.7}
\definecolor{color2}{rgb}{0.85,0,0}

\usepackage[breaklinks=true,colorlinks,citecolor=red,linkcolor=color2,urlcolor=color1]{hyperref}

\newcommand{\eref}[1]{\textcolor{color2}{\hyperref[#1]{eq.~(\ref{#1})}}}
\newcommand{\Eref}[1]{\textcolor{color2}{\hyperref[#1]{Eq.~(\ref{#1})}}}
\newcommand{\fref}[1]{\textcolor{color2}{\hyperref[#1]{Fig.~\bfseries{\ref{#1}}}}}
\newcommand{\sref}[1]{\textcolor{color2}{\hyperref[#1]{Sec.~\bfseries\ref{#1}}}}

\newcommand{\aref}[1]{\textcolor{color2}{\hyperref[#1]{App.~\bfseries\ref{#1}}}}
\newcommand{\tr}{{\rm Tr}}

\usepackage{titlesec}
\titlespacing*{\section}{0pt}{10pt plus 4pt minus 5pt}{7pt plus 3pt minus 5pt}
\titlespacing*{\subsection}{0pt}{5pt plus 3pt minus 1pt}{4pt plus 2pt minus 4pt}
\titleformat{\section}{\centering\bfseries}{\thesection.}{.5em}{}

\begin{document}

\title{Quantum Suicide in Many-Worlds Implies P=NP}

\author{Veronika Baumann}
\altaffiliation{These authors are equally to blame for this work.}
\affiliation{Institute for Quantum Optics and Quantum Information - IQOQI Vienna, Austrian Academy of Sciences, Boltzmanngasse 3, A-1090 Vienna, Austria}
\affiliation{Atominstitut, TU Wien, 1020 Vienna, Austria}

\author{Alberto Rolandi}
\altaffiliation{These authors are equally to blame for this work.}
\affiliation{Atominstitut, TU Wien, 1020 Vienna, Austria}
\affiliation{Institute for Quantum Optics and Quantum Information - IQOQI Vienna, Austrian Academy of Sciences, Boltzmanngasse 3, A-1090 Vienna, Austria}

\date{April 1, 2026}

\begin{abstract}
\noindent In this paper we propose a totally serious~\cite{Skotiniotis20,Winter21,Winter24} algorithm to solve NP problems in polynomial time provided one is willing to wager the fate of all observers in the universe on the many-world interpretation of quantum theory being correct.
\end{abstract}

\maketitle

\section{Introduction}

The Millennium Prize problem \emph{P versus NP}~\cite{Cook1971,Levin1973,Papadimitriou1994,Cook2006} is one of the current major open questions in computer science. It addresses the question of whether solving a problem is fundamentally harder than verifying the correctness of a proposed solution. ``P'' and ``NP'' refer to computational complexity classes in classical computing. If the time required\footnote{Measured in number of basic operations.} to solve a problem scales polynomially with the input size then that question belongs to the class of problems $P$. However, for most problems there is no known algorithm solving them in polynomial time. For some of these, it is instead possible to verify quickly, i.e. in a time that scales polynomially with the input size. If a proposed solution to the problem is correct; this class of problems is known as \emph{non-deterministic polynomial time}~\cite{Arora2009}, or $NP$ for short. The $P$ versus $NP$ problem asks whether these two sets of problems coincide or are different. It is quite direct to see that $P\subseteq NP$, as one can verify any proposed solution to a problem in $P$ by solving said problem and comparing the actual to the proposed solution~\cite{Cormen2009}. It might seem intuitive that these two classes are different, since verifying a solution is generally easier than finding one. For example, for a non-prime integer $N$ it is difficult to find its prime decomposition $N = p_1^{k_1}\dots p_n^{k_n}$. However, given a candidate decomposition $\{(p_1,k_1),\dots,(p_n,k_n)\}$ it is very easy to check if we recover $N$ from it by simply multiplying these numbers together. Despite the intuitive aspect of this problem, there is no example of an $NP$ problem that is \emph{proven to not belong} to $P$.

In general, while it is not known whether $NP$ problems can be solved in polynomial time, one can always obtain a solution in exponential time. To achieve this scaling, one simply tests candidate solutions one by one using a polynomial-time verification algorithm. Since the space of possible solutions scales exponentially with the input size and the verification time is polynomial (by definition of $NP$), it takes on average an exponential time to find the solution of any $NP$ problem using this algorithm. However, and crucially for this letter, it is possible that the correct solution happens to be the first candidate tested. In general, it is exponentially unlikely to make such a lucky guess. In this work we argue that, by exploiting the many-worlds interpretation of quantum theory, one can effectively arrange to “get lucky’’ every time such an algorithm is run.

\begin{figure}[!tb]
	\centering
	\includegraphics[width=\columnwidth]{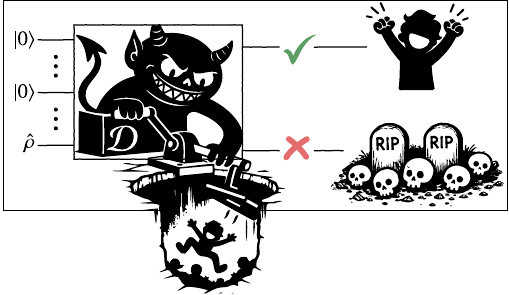}
    \vspace{-15pt}
	\caption{Depiction of a possible implementation of the proposed algorithm, where the \emph{Doomsday Channel} (cf. \sref{sec:D}) is realized by Maxwell's demon.}
    \vspace{-15pt}
	\label{fig:class_fals}
\end{figure}

The \emph{quantum suicide} thought experiment~\cite{squires1994mystery,lewis2000like,lewis2004many} extends the famous Schrödinger's cat to human experimenters and explores its implication in a many worlds interpretation of quantum theory~\cite{vaidman2001probability,wallace2002worlds,deutsch2011beginning}. Imagine an experimenter and a quantum system in an isolated box and a mechanism by which one state of the quantum system leads to the death of the experimenter, while the orthogonal state leaves them alive. Analogous to Schrödinger's cat this leads to a quantum state
\begin{equation}
\alpha \ket{0}_S\ket{\text{``dead''}}_E +\beta \ket{1}_S\ket{\text{``alive''}}_E ,
\label{cat-state}
\end{equation}
where $S$ and $E$ stand for the system and experimenter respectively and $\alpha,\beta$ are complex numbers determined by the initial state $\ket{\psi}= \alpha \ket{0}+\beta \ket{1}$ of $S$. Once other experimenters open the box, they will find $E$ dead with probability $|\alpha|^2$ and alive with $|\beta|^2$. Different interpretations of quantum theory give different accounts of the state in \eref{cat-state}. The many-worlds interpretation, however, maintains that this state describes two alternative branches of reality that both exist, one in which $E$ is dead and one in which $E$ is alive. The probabilities for these two alternatives are the same as before, meaning that when other experimenters open the box they can expect to find $E$ dead with probability $|\alpha|^2$ and alive with $|\beta|^2$.
The state $\ket{\text{``dead''}}$ signifies that experimenter $E$ has no more experience of this branch of the wave function of the universe\footnote{While seemingly straightforward, this assertion is actually contingent on there not being an afterlife and hence death ending all possible experience~\cite{Dead,Zoro,Ovid08}.}. Hence, subjectively they can only end up in the $\ket{\text{``alive''}}$ state, which amounts to post selecting on this particular branch of the wave function of the universe.
Unless $\beta=0$, there will be a world, in which $E$ is still alive also for multiple repetitions of the experiment. Subjectively, $E$ will ``always'' survive and find the system $S$ in state $\ket{0}$, which has been called quantum immortality~\cite{tegmark1998interpretation}.

\section{The Doomsday Algorithm}\label{sec:D}

In this section we propose a modification of the original quantum suicide experiment, which would allow humanity to solve $NP$ problems in polynomial time. The main idea relies on exploiting that the verification of a proposed solution to an $NP$ problem can be done in polynomial time. By proposing solutions at random, it is very unlikely to solve the problem in less than exponential time. However, in the quantum suicide thought experiment it does not matter how unlikely the \emph{alive outcome} is (as long as its probability is non-zero). Originally, this outcome has no particular significance, as it is the result of a measurement performed on an arbitrarily prepared quantum system. Here instead, we choose the state being measured as the output of a quantum algorithm. By making this a verification algorithm of an $NP$ problem, the agent would either obtain the solution of the problem in polynomial time (or perish). For the agent, running this experiment would create the subjective impression that ``$P=NP$''. However, to any outside observer, the agent would almost always end up dead. Yet, by extending the experiment to all observers it would lead to an inter-subjective assertion ``$P=NP$'' that is indistinguishable from the objective statement $P=NP$.\\

The entire procedure can be expressed as a quantum algorithm, which is as follows.
\hspace*{-0.2cm}
\begin{quantikz}[wire types={c,q,n,q,q,b},column sep=0.75cm,row sep=0.4cm]
    \lstick{$\lambda$}&[-0.3cm]\slice{A}&\ctrl{1}\slice{B}&\slice{C}&&\\
    \lstick{\ket{0}}&\gate{H}&\gate[4][1.7cm]{U}\gateinput[3]{$s$}\gateoutput[3]{$s$}&&\meter{}&\setwiretype{c}\rstick[wires=3]{\text{output}}\\
    \vdots&\vdots&&\vdots&\vdots\\
    \lstick{\ket{0}}&\gate{H}&&&\meter{}&\setwiretype{c}\\
    \lstick{\ket{0}}&&\gateinput{$y$}\gateoutput{$y\oplus f(s)$}&\octrl{1}&&\\
    \lstick{$\hat\rho$}&&&\gate{\mathcal D}&&
\end{quantikz}
The input state of the verification algorithm is $\ket{\psi^0} = \ket{0^n}_S\otimes\ket{0}_A$, where $S$ will encode the possible solutions to the NP problem, the ancilla $A$ will encode if the corresponding bit-string is a solution (in which case it will be in state $\ket{1}$). Moreover, $\hat\rho$ is the state of all observers, obtained from the wavefunction of the entire universe, and the classical variable $\lambda$ sets the parameters of the NP problem being solved. After the first set of unitaries, the state of $S$ is given by
\begin{equation}
    \ket{\psi^A}_S = \frac{1}{2^n}\sum_{s\in \mathcal B_n}\ket{s}_S,
\end{equation}
where $\mathcal B_n$ is the set of all bit strings of length $n$. Therefore the total state at slice A is given by
\begin{equation}
    \hat\sigma_A = \ket{\psi^A}\!\bra{\psi^A}_S \otimes \ket{0}\!\bra{0}_A\otimes\hat \rho.
\end{equation}

The \emph{verification unitary} $U$ checks whether a bit string $s$ is a solution to a certain problem. By definition of $NP$
it has polynomial complexity.  The unitary acts on $S$ and $A$ in such a way that it will encode in $A$ if the state of $S$ is a solution of the problem. More precisely, it is defined as 
\begin{equation}
U \ket{s, y} = \ket{s, y \oplus f(s)} \, ,
\label{eq:U_solution}
\end{equation}
where $f(s)= 1$ if $s$ is a solution and $f(s)= 0$ otherwise. Hence, after the application of this unitary we arrive at an overall state 
\begin{equation}
    \hat\sigma_B = \ket{\psi^B}\!\bra{\psi^B}_{SA}\otimes\hat \rho
\end{equation}
at slice B, with
\begin{equation}
    \ket{\psi^B}_{SA} =\frac{1}{2^n}\left(\ket{s^*}_S\otimes\ket{1}_A + \!\!\sum_{s \in \mathcal B_n\setminus s^*}\!\! \ket{s}_S\otimes\ket{0}_A\right),
\end{equation}
where $s^*$ is the solution to the problem.\\

The \emph{doomsday channel} $\mathcal D$ is an operation on the quantum state of all observers: it kills them. For example, we can suppose it brings them to their global thermal state $\hat\omega_\beta = \frac{e^{-\beta \hat H}}{\tr[e^{-\beta H}]}$, for $\hat H$ the Hamiltonian of all observers and $\beta$ some arbitrary inverse temperature. By the simplicity of this operation and for morality reasons, we assume that it can be done in $\mathcal O(1)$ time. In this algorithm, it is (anti-)controlled by the ancilla $A$. Therefore if the ancilla is in state $\ket{1}$ it is not applied, and it is applied (thus killing every single observer) if $A$ is in state $\ket{0}$.\footnote{One could potentially have the doomsday channel destroy the rest of the universe as well instead of just all observers and the result would be the same. However, we felt that in this case the machine would have been too powerful and unrealistic.} This means, that at slice C we arrive at the final state
\begin{align}
    \hat\sigma_C =& \frac{1}{2^n}\Big(\ket{s^*}\!\bra{s^*}_S\otimes\ket{1}\!\bra{1}_A \otimes \hat \rho  \\
    &+ \!\!\sum_{s \in \mathcal B_n\setminus s^*}\!\! \ket{s}\!\bra{s^*}_S\otimes\ket{0}\!\bra{0}_A\otimes \hat\omega_\beta \Big),
\end{align}
which when post selected on there existing observers in the universe is just
\begin{equation}
    \hat\sigma_C^{\rm alive} = \ket{s^*}\!\bra{s^*}_S\otimes\ket{1}\!\bra{1}_A \otimes \hat\rho.
\end{equation}
The solution to the NP problem is then obtained by measuring the qubits of $S$ in the computational basis. This means that at the end, in the branch of the wavefunction of the universe which still contains observers, said observers have solved an NP problem in polynomial time.

\section{Conclusion}
In this paper we described an algorithm analogous to a ``lucky'' brute-force search, where the first guess happens to be the correct solution. By exploiting the post-selection inherent to the quantum suicide thought experiment, we ensure that surviving observers are ``lucky'' at every implementation. They would therefore unanimously conclude that $NP$ problems can be solved in polynomial time — a consensus indistinguishable, to them, from the objective statement $P=NP$.

The price of this \emph{quantum advantage}, however, is the annihilation of all observers in the remaining $2^n - 1$ branches of the universe. The exponential overhead is not eliminated but merely displaced: from runtime to body count. We therefore conclude that any $NP$ search problem can be solved efficiently, provided one is willing to threaten the existence of all observers\footnote{Potentially other observers beyond humanity as well.}---and, crucially, that at least one branch survives to write up the results.

\section*{Acknowledgements}
We acknowledge Alê for producing Fig. 1, as part of his ongoing efforts to establish himself as the world’s leading Maxwell’s demon designer. We further thank Tom Rivlin for his input on humour and creative writing. We promise this paper did not involve any use of AI. Only humans would be deranged enough to write what we did here.

\newpage

\bibliography{refs}

@book{squires1994mystery,
  title={The mystery of the quantum world},
  author={Squires, Euan J},
  year={1994},
  publisher={CRC Press}
}

@article{wallace2002worlds,
  title={Worlds in the Everett interpretation},
  author={Wallace, David},
  journal={Studies in History and Philosophy of Science Part B: Studies in History and Philosophy of Modern Physics},
  volume={33},
  number={4},
  pages={637--661},
  year={2002},
  publisher={Elsevier}
}

@book{Dead,
    author = {Priests of ancient Egypt over 1000 years},
    title = {Book of the Dead},
    publisher = {Pharaon's Publishing Services, Memphis, Inebu-hedj},
    year = {ca. 1550 BCE},
    url = {https://collezioni.museoegizio.it/en-GB/material/S_8438},
}

@book{Zoro,
    author =  {Zarathushtra Spitama},
    title = {Avesta},
    publisher = {Sasanian Publishing Services, Istakhr, Sassanian Empire},
    year = {ca. 1323 BCE},
    url = {https://en.wikipedia.org/wiki/Zoroastrianism}
}

@book{Ovid08,
    author = {Publius Ovidius Naso},
    title = {Metamorphosis},
    publisher = {Emperor's Publishing Services, Rome, Roman Empire},
    year = {8 CE}
}

@book{Arora2009,
  title = {Computational Complexity: A Modern Approach},
  ISBN = {9780511804090},
  url = {http://dx.doi.org/10.1017/CBO9780511804090},
  DOI = {10.1017/cbo9780511804090},
  publisher = {Cambridge University Press},
  author = {Arora,  Sanjeev and Barak,  Boaz},
  year = {2009},
  month = apr 
}

@misc{Winter24,
  doi = {10.48550/ARXIV.2403.19977},
  url = {https://arxiv.org/abs/2403.19977},
  author = {Winter,  A. and Winter,  A. and Winter,  A. and Winter,  A.},
  title = {Is Winter Coming?},
  publisher = {arXiv},
  year = {2024},
  copyright = {Creative Commons Attribution 4.0 International}
}

@misc{Winter21,
  doi = {10.48550/ARXIV.2103.16662},
  url = {https://arxiv.org/abs/2103.16662},
  author = {Winter,  Andreas},
  title = {A genuinely natural information measure},
  publisher = {arXiv},
  year = {2021},
  copyright = {Creative Commons Attribution 4.0 International}
}

@misc{Skotiniotis20,
  doi = {10.48550/ARXIV.2003.13715},
  url = {https://arxiv.org/abs/2003.13715},
  author = {Skotiniotis,  Michalis and Winter,  Andreas},
  title = {Quantum Godwin's Law},
  publisher = {arXiv},
  year = {2020},
  copyright = {arXiv.org perpetual,  non-exclusive license}
}

@inproceedings{Papadimitriou1994,
  series = {SCT-94},
  title = {The complexity of optimal queueing network control},
  url = {http://dx.doi.org/10.1109/SCT.1994.315792},
  DOI = {10.1109/sct.1994.315792},
  booktitle = {Proceedings of IEEE 9th Annual Conference on Structure in Complexity Theory},
  publisher = {IEEE Comput. Soc. Press},
  author = {Papadimitriou,  C.H. and Tsitsiklis,  J.N.},
  pages = {318–322},
  collection = {SCT-94},
  year = {1994}
}

@BOOK{Cormen2009,
  title     = "Introduction to Algorithms",
  author    = "Cormen, Thomas H and Leiserson, Charles E and Rivest, Ronald L and Stein, Clifford",
  publisher = "MIT Press",
  series    = "The MIT Press",
  edition   =  3,
  month     =  jul,
  year      =  2009,
  address   = "London, England"
}

@book{Cook2006,
  author    = {Cook, Stephen},
  title     = {The P versus NP Problem},
  publisher = {Clay Mathematics Institute},
  year      = {2006},
  series    = {Clay Mathematics Institute Millennium Prize Problems}
}

@inproceedings{Cook1971,
  series = {STOC ’71},
  title = {The complexity of theorem-proving procedures},
  url = {http://dx.doi.org/10.1145/800157.805047},
  DOI = {10.1145/800157.805047},
  booktitle = {Proceedings of the third annual ACM symposium on Theory of computing  - STOC ’71},
  publisher = {ACM Press},
  author = {Cook,  Stephen A.},
  year = {1971},
  pages = {151–158},
  collection = {STOC ’71}
}

@article{Levin1973,
  author       = {Levin, Leonid A.},
  title        = {Universal Search Problems},
  journal      = {Problems of Information Transmission},
  year         = {1973},
  volume       = {9},
  number       = {3},
  pages        = {265--266},
  note         = {Translated from \emph{Problemy Peredachi Informatsii} 9(3):115--116 (1973)}
}

@article{lewis2000like,
  title={What is it like to be Schr{\"o}dinger's cat?},
  author={Lewis, Peter J},
  journal={Analysis},
  pages={22--29},
  year={2000},
  publisher={JSTOR}
}

@article{lewis2004many,
  title={How many lives has Schr{\"o}dinger's cat?},
  author={Lewis, David},
  journal={Australasian Journal of Philosophy},
  volume={82},
  number={1},
  pages={3--22},
  year={2004},
  publisher={Taylor \& Francis}
}

@article{tegmark1998interpretation,
  title={The interpretation of quantum mechanics: Many worlds or many words?},
  author={Tegmark, Max},
  journal={Fortschritte der Physik: Progress of Physics},
  volume={46},
  number={6-8},
  pages={855--862},
  year={1998},
  publisher={Wiley Online Library}
}

@book{deutsch2011beginning,
  title={The beginning of infinity: Explanations that transform the world},
  author={Deutsch, David},
  year={2011},
  publisher={penguin uK}
}

@article{vaidman2001probability,
  title={Probability and the Many-Worlds interpretation of quantum theory},
  author={Vaidman, Lev},
  journal={arXiv preprint quant-ph/0111072},
  year={2001}
}
\appendix

\widetext

\end{document}